\documentclass[12pt]{iopart}
\usepackage{amsmath}
\usepackage{iopams}

\begin{document}

\maketitle

\title{Quantum Theory of Cold Dark Matter Halos}

\author{Z. E. Musielak}
\address{Department of Physics, The University of Texas at 
Arlington, Arlington, TX 76019, USA}
%

\begin{abstract}
A nonrelativistic quantum theory of dark matter particles 
in a spherical halo is developed by using a new asymmetric 
equation, which is complementary to the Schr\"odinger
equation.  The theory predicts that each dark matter 
halo has its core and envelope with very distinct physical 
properties.  The core is free of any quantum structure and 
its dark matter particles are in random motion and frequently 
collide with each other.  However, the envelope has a 
global quantum structure that contains quantized orbits 
populated by the particles.  Applications of the theory to 
dark matter halos with given density profiles are described, 
and physical implications and predictions of the theory are 
discussed. 
\end{abstract}


\section{Introduction}

Many theories of dark matter (DM) have been proposed [1-20].
One popular theory predicts the existence of weakly interacting 
massive particles (WIMPs) of dark matter (DM). However, so far, 
all attempts to discover WIMPs experimentally have failed (e.g., 
[21-24]); thus, the nature and origin of DM still remains unknown.  
According to the Planck 2018 mission [25], DM constitutes 26.8\% 
of the total mass-energy density of the Universe, which is almost 
5.5 times more than the amount of ordinary matter (OM). 

Recent results [19] show that the Schr\"odinger equation (SE) 
of nonrelativistic quantum mechanics [26] is consistent with 
the irreducible representations (irreps) of the extended 
Galilean group [27-31].  The validity of the SE and its 
theoretical predictions have been verified by numerous 
experiments performed since Schr\"odinger published it 
[32].  The obtained theoretical and experimental results 
have demonstrated that the SE describes OM and its 
quantum properties.  Thus, it seems natural to apply the 
SE to construct models of DM, which require solving the 
SE on large galactic scales [33,34].  However, more 
detailed calculations [35,36] have revealed that such 
models constructed for different galactic halos require 
DM particles to have different masses, which is difficult 
to justify physically. 

The irreps used to obtain the SE also allow for the 
existence of a new asymmetric equation (NAE) that is 
complementary to the SE [19]; note that both the SE 
and NAE can be used to describe propagation of 
classical waves [37].  Moreover, since all models of 
DM based on the SE have failed [35,36], an attempt 
was made to construct a model of two DM particles 
interacting only gravitationally by using the NAE [20].  
The model has two main shortcomings, namely, it does 
not consider any halo's effects on the particles, and it is 
based on the assumption that the density is constant.  In 
this paper, a quantum theory of a DM halo is developed 
based on the NAE; the main advantage of the theory is 
that it removes all the shortcomings of the previous 
model. 

The developed theory predicts a quantum structure of the 
halo that resembles an atom; means that each halo has 
a core and an envelope that surrounds it.  The core 
occupies a small fraction ($< 10\%$) of the volume of 
the halo, and its density is constant, which makes the 
core free of any quantum structure.  As a result, the DM 
particles move randomly inside the core and frequently 
collide with each other.  On the other hand, the envelope
that surrounds the core contains quantized orbits with DM
particles confined to them.  A narrow region of the envelope
in the vicinity of the core is of special interest as it allows 
for exchanging DM particles and quanta of energy (dark 
gravitons) between the core and the envelope.  Applications 
of the theory to DM halos with constant density and with 
density varying linearly with halo's radius are presented, 
and the obtained results are discussed.

This paper is organized as follows: nonrelativistic asymmetric 
equations are derived in Section 2; previous theories of DM 
and their validity are discussed in Section 3; a quantum theory 
of DM particles confined to a spherical DM halo is formulated in 
Section 4;  physical implications of the theory and its predictions 
are discussed in Section 5; and conclusions are given in Section 6.   

\section{Nonrelativistic asymmetric equations}

\subsection{Group theory derivation of asymmetric equations}
 
The structure of the extended Galilean group is $\mathcal {G}_e 
= [O(3) \otimes_s B(3)] \otimes_s [T(3+1) \otimes U(1)]$, 
where $O(3)$ and $B(3)$ are subgroups of rotations and boosts, 
$T(3+1)$ is an invariant subgroup of combined translations in 
space and time, and $U(1)$ is a one-parameter unitary subgroup;
the irreps of $\mathcal {G}_e$ are well-known [27-31].  

A scalar function $\phi (t, \mathbf {x})$ transforms as one of 
the irreps of $\mathcal {G}_e$ if, and only if, the following 
eigenvalue equations $i \partial_t \phi = \omega \phi$ and $- i 
\nabla \phi = \mathbf {k} \phi$ are satisfied, where $\partial_t =
\partial /\partial t$, and $\omega$ and $\mathbf {k}$ are labels 
of the irreps. The eigenvalue equations can be used to derive 
the following two second-order asymmetric equations [19,20]
\begin{equation}
\left [ i {{\partial} \over {\partial t}} + \left ( \frac{\omega}{k^2} 
\right ) \nabla^{2} \right ] \phi (t, \mathbf {x}) = 0\ ,
\label{eq1}
\end{equation}  
and 
\begin{equation}
\left [ {{\partial^2} \over {\partial t^2}} -  i \left ( \frac{\omega}{k} 
\right )^2 {\mathbf k} \cdot \nabla \right ] \phi (t, \mathbf {x}) = 0\ .
\label{eq2}
\end{equation}  
where the ratios of the labels of the irreps in both equations are 
constant as required by the eigenvalue equations.  Comparison 
of these equations shows that their mathematical structure is 
complementary, but their wavefunctions must be different.  

The ratios of eigenvalues in the above equations can be expressed 
in terms of the universal constants, so the equations can be used 
to formulate quantum theories of OM and DM [19,20].  However, 
if the ratios are expressed in terms of the wave characteristic speed 
and frequency, then the equations become the wave equations that 
describe propagation of classical waves in uniform or nonuniform 
media [36]; classical waves are not discussed in this paper.

\subsection{The Schr\"odinger equation}

By using the de Broglie relationship [26], the ratio of eigenvalues 
in Eq. (\ref{eq1}) can be evaluated as $\omega / k^2 = \hbar / 
2m$, which turns the equation into the Schr\"odinger equation 
[32]. Denoting the wavefunction by $\phi_S$, the SE can be 
written in its standard form [26] as
\begin{equation}
\left [ i {{\partial} \over {\partial t}} +  \frac{\hbar}{2 m}
\nabla^{2} \right ] \phi_S (t, \mathbf {x}) = 0\ ,
\label{eq3}
\end{equation}  
where $m$ is the mass of a particle.

The eigenvalue equations can also be used to define operators
of energy, $\hat E$, and momentum, $\mathbf {\hat P}$. This
is achieved by multiplying the eigenvalue equations by $\hbar$, 
and defining $\hat E = i \hbar \partial_t$ and $\mathbf {\hat P} 
= - i \hbar \nabla$.  Then, the eigenvalue equations for these 
operators are: $\hat E \phi_S = E \phi_S$ and $\mathbf {\hat P} 
\phi_S = {\mathbf p} \phi_S$, with the eigenvalues $E = \hbar 
\omega$ and $\mathbf {p} = \hbar \mathbf k$.  Using these 
results, the SE can be written in its operator's form as
\begin{equation}
\hat E \phi_S = \frac{1}{2m} \left ( \mathbf {\hat P} \cdot 
\mathbf {\hat P} \right ) \phi_S\ , 
\label{eq4}
\end{equation}  
which gives the following relationship between the eigenvalues 
\begin{equation}
E = \frac{\mathbf {p} \cdot \mathbf {p}}{2 m} = 
\frac{p^2}{2 m} =  E_{k} \equiv E_{SE}\ .  
\label{eq5}
\end{equation}  
This shows that the SE is based on the nonrelativistic kinetic 
energy, $E_{k} = E_{SE}$, which is a well-known result 
(e.g., [26]). 

\subsection{A new asymmetric equation}

The de Broglie relationship [26] may also be used to determine the 
ratio of eigenfunctions $\omega^2 / k^2 = \varepsilon_o / 2m$, 
where $\varepsilon_o = \hbar \omega_o$ is a fixed quanta of 
energy, and $\omega_o$ is a fixed frequency [19,20].  Then, 
Eq. (\ref{eq2}) becomes
\begin{equation}
\left [ {{\partial^2} \over {\partial t^2}} -  i \frac{\varepsilon_o}
{2 m}{\mathbf k} \cdot \nabla \right ] \phi_A (t, \mathbf {x}) = 0\ ,
\label{eq6}
\end{equation}  
where $\phi_A (t, \mathbf {x})$ is the wavefunction of this new
asymmetric equation (NAE).  

To determine the expression for energy that underlies the NAE,
the eigenvalue equations for the operators of energy, $\hat E$, 
and momentum, $\mathbf {\hat P}$ must be determined. Since
the wavefunctions of the NAE and SE are different, the eigenvalue
equations for $\phi_A$ will be different than those found for 
$\phi_S$ in Section II.B.  The main reason is that the operators 
$\hat E$ and $\mathbf {\hat P}$ acting on different eigenfunctions 
give different eigenvalues. 

Thus, for the energy, $\hat E = i \hbar \partial_t$, and momentum,
$\mathbf {\hat P} = - i \hbar \nabla$, operators, the eigenvalue 
equations are: $\hat E \phi_A = \sqrt{\varepsilon_o E}\ \phi_A$ 
and $(\hbar \mathbf {k} \cdot \mathbf {\hat P}) \phi_A =(\mathbf 
{p} \cdot \mathbf {\hat P}) \phi_A = p^2 \phi_A$, with the 
eigenvalues being $\varepsilon_o = \hbar \omega_o$, $E = 
\hbar \omega$ and $\mathbf {p} = \hbar \mathbf {k}$.  Then, 
Eq. (\ref{eq6}) can be written in the following form
\begin{equation}
\left ( \frac{1}{\varepsilon_o} \right ) \hat E^2 \phi_A = 
\frac{1}{2m} \left ( \mathbf {p} \cdot \mathbf {\hat P} 
\right ) \phi_A\ , 
\label{eq7}
\end{equation}  
and the relationship between the eigenvalues becomes
\begin{equation}
E = \frac{p^2}{2 m} = E_{k} \equiv E_{NAE}\ , 
\label{eq8}
\end{equation}  
which shows that $E_{NAE} = E_{SE} = E_{k}$ and, 
as expected, both the NAE and SE are based on the 
nonrelativistic kinetic energy.

The presented results demonstrate that the main difference 
between the SE and NAE is that the former allows for the 
quanta of energy $\hbar \omega$ to be of any frequency; 
however, the latter is valid only when the quanta of energy 
$\varepsilon_o$ is fixed at one frequency $\omega_o$.  
Moreover, the evolution of the wavefunction $\phi_A (t, 
\mathbf {x})$ described by the NAE depends on direction
as shown by the presence of the term ${\mathbf k} \cdot 
\nabla \phi$, whose values are different in different directions 
with respect to ${\mathbf k}$; the SE does not show such 
a directional dependence. 

\section{Asymmetric equations and quantum theories of dark matter}

The fact that the SE describes the quantum structure of OM
has been known for almost 100 years (e.g., [26]). There 
were also attempts to use the SE to formulate quantum 
theories of DM. The basic idea was proposed by Sin [33], 
who postulated the existence of extremely light bosonic 
DM particles with masses of the order of $10^{-24}$ eV, 
which allows for solving the SE on the galactic scale 
because of the very long Compton wavelength of such 
particles. The gravitational potential added to the SE was 
calculated by solving the Poisson equation with the DM 
density as the forcing term.  The work was followed by 
Hu et al. [34], whose DM particles had masses of the 
order of $10^{-22}$ eV.  However, more detailed 
studies [35,36] revealed that these theories require 
different masses of DM particles in different galactic 
halos, which is difficult to justify from a physical point 
of view. 
    
Since the SE equation describes the quantum structure 
of OM, and since the developed quantum theories of 
DM based on the SE failed, it was suggested that the 
NAE, being complementary to SE, may represent the 
quantum structure of DM [19,20].  If this is correct, 
then DM particles may only exchange the quanta of 
energy $\varepsilon_o = \hbar \omega_o$, whose 
frequency $\omega_o$ is fixed for DM particles.  
This means that while OM emits or absorbs radiation 
at a broad range of frequencies, DM's emission or 
absorption is restricted to only one specific frequency 
$\omega_o$ that is characteristic for DM particles. 

An attempt to formulate a nonrelativistic quantum 
theory of cold DM based on the NAE was done in 
[20].  In this theory, a pair of DM particles interact
only gravitationally and the particles are represented
by a scalar wavefunction, whose evolution in time 
and space is described by the NAE. The obtained 
solutions for the wavefunction show that the particles
may exchange the quanta of energy $\varepsilon_o$ 
between themselves or with other particles in the 
halo.  Moreover, the velocities of the particles in 
the pair may change due collissions with other 
DM particles.  

The main disadvantage of the theory developed in 
[20] is that all gravitational effects of the halo on 
the pair of DM particles were neglected, and that 
the density in the halo was assumed to be constant, 
which is not supported by the known halo models 
(e.g., [38-41]).  In the following, the theory is 
generalized to account for the effects of the halo
including non-constant density.  

\section{Dark matter halos and their quantum structure}

\subsection{Quantum theory of halos}

Consider a spherical halo of DM particles with radius $R_{h}$ 
and total mass $M_{h}$.  Let a DM particle of mass $m$ 
be located at point $P$, whose distance from the center 
of the halo is $R$ (spherical coordinate).  The force acting 
on the particle is $F_g (R) = G M(R)\ m / R^2$, and this 
force is balanced by the centrifugal force $F_c (R) = m R\ 
\Omega_c^2 (R)$, which gives the circular orbital frequency 
of the particle 
\begin{equation}
\Omega_c^2 (R) = \frac{G M( R)}{R^3}\ .
\label{eq9}
\end{equation}  
where $M (R) = 4 \pi \int_0^R \rho (\tilde R) \tilde {R}^2 
d \tilde {R}$
with $\rho (R)$ being the density profile of DM inside the halo,
which must be specified.  

The new asymmetric equation given by Eq. (\ref{eq6})
describes the temporal and spatial evolution of the wavefunction
$\phi_A (t, R)$ for free DM particles.  In a galactic DM halo, 
the particles are affected by the halo's gravitational field and
undergo gravitational interactions between themselves; in 
addition, the DM particles may also collide.  In the theory 
developed in this paper, the effects of halo's gravitational 
field on the particles are fully accounted for.  However, 
gravitational interactions between two DM particles are 
neglected because they are very weak [20].  Moreover, 
as the obtained results show, collisions are only important 
in the cores and their roles in halo's envelopes are negligible.  
This shows that the NAE for free particles must be modified 
to take into account interactions of DM particles with the 
halo's gravitational field.  To achieve it, $\Omega_{c}^2 (R)$
is added to Eq. (\ref{eq6}) as it plays the role of gravitational 
potential.

The resulting equation
\begin{equation}
\left [ {{\partial^2} \over {\partial t^2}} -  i \frac{\varepsilon_o}
{2m} \vert \mathbf k \cdot \mathbf {\hat R} \vert \frac {\partial}
{\partial R} + \Omega^2_c (R) \right ] \phi_A (t, R) = 0\ ,
\label{eq10}
\end{equation}  
becomes the governing equation for the theory 
developed in this paper.  After separating the variables, 
$\phi_A (t, R) = \chi (t) \eta (R)$, the equation for 
$\eta (R)$ is 
\begin{equation}
\frac{d \eta}{\eta} = i \frac{2m}{\varepsilon_o \vert \mathbf k 
\cdot \mathbf {\hat R} \vert} \left [ \mu^2 - \Omega^2_c (R) 
\right ] dR\ ,
\label{eq11}
\end{equation}  
where $\mathbf {\hat R} = \mathbf {R} / R$, and $\mu^2$ is 
the separation constant to be determined.  

Then, the real part of the solution of Eq. (\ref{eq11}) is   
\begin{equation}
\eta (R) = \eta_o \cos \left ( \frac{2 G m^2 [ \mu^2 R - I_c (R)]} 
{\varepsilon_o^2 \vert \mathbf {\hat k} \cdot \mathbf {\hat R} 
\vert } \right )\ ,
\label{eq12}
\end{equation}  
where $I_c (R) = \int_0^R \Omega_c^2 (\tilde R) d \tilde {R}$, and 
$\mathbf {k} = k \mathbf {\hat k}$, with $k = 1 / \lambda_o = 
G m^2 / \varepsilon_o$ = const [20]. 

The maxima of the cos function are when its argument is $\pm 2 n \pi$,
with $n = 0$, 1, 2, 3, ... .  After using $\varepsilon_o = \hbar
\omega_o$, the condition for the maxima can be written as 
\begin{equation}
\mu^2 - \frac{1}{R} I_c (R)  = \pm n \pi \omega_o^2 \vert \mathbf 
{\hat k} \cdot \mathbf {\hat R} \vert \frac{\hbar^2}{G m^3 R}\ .
\label{eq13}
\end{equation}  

To determine the separation constant $\mu^2$, the integral $I_c (R)$ 
must be evaluated in the entire halo that is from $R = 0$ to $R = R_h$.
It can be shown that the result of this integration is  
\begin{equation}
I_c (R_h) = \int_0^{R_h} \Omega_c^2 (\tilde R) d \tilde {R}
= C_{\rho} R_h \Omega_h^2\ ,
\label{eq14}
\end{equation}  
where $C_{\rho}$ is a dimensionless constant whose value
is different for different density profiles (see specific examples
in Sections 4.2 and 4.3).  Then, the separation constant 
$\mu^2$ is given by 
\begin{equation}
\mu^2 = C_{\rho} \Omega_h^2 \pm n \pi \omega_o^2 \vert 
\mathbf {\hat k} \cdot \mathbf {\hat R_h} \vert \frac{\hbar^2}
{G m^3 R_h}\ .
\label{eq16}
\end{equation}  

Taking $\mu^2 = C_{\rho} \Omega_n^2$, with $\Omega_n$ being 
the quantized orbital frequencies, Eq. (\ref{eq16}) becomes
\begin{equation}
\Omega_n^2 = \Omega_h^2 \pm n \pi \kappa_h \omega_o^2\ ,
\label{eq17}
\end{equation}  
where $+$ and $-$ is for the inside and outside of the halo, 
respectively, and $\kappa_h = \hbar^2 / (G m^3 C_{\rho} R_h)$ 
is the dimensional constant that connects the universal constants 
$G$ and $\hbar$ and the mass $m$ of DM particles to the halo 
radius $R_h$ and the density gradient constant $C_{\rho}$.  
Moreover, the term $\vert \mathbf {\hat k} \cdot \mathbf {\hat R_h}
\vert = 1$ because the unit vector $\mathbf {\hat R_h}$ is 
not restricted and thus it can always be aligned with the unit 
vector $\mathbf {\hat k}$.  Multiplying Eq. (\ref{eq17}) by 
$\hbar^2$, the spectrum of quantum energies corresponding 
to the orbits $\Omega_n^2$ inside ('+') and outside ('-') the 
halo is obtained 
\begin{equation}
E_n^2 = E_h^2 \pm n \pi \kappa_h \varepsilon_o^2\ .
\label{eq18}
\end{equation}  

To calculate the radius of the quantized orbits of the halo, the 
conservation of angular momentum, $L_n = m R_n^2 \Omega_n 
=$ const, must be used.  Since the value of $L_0$ is fixed and 
known, $L_0 = m R_0^2 \Omega_0 = m R_h^2 \Omega_h =  
L_h = $ const, the radius $R_n$ is obtained from $L_n = L_h$ 
with $n\geq 0$, which gives 
\begin{equation}
R_n = R_h \sqrt{\frac{\Omega_h}{\Omega_n}}\ ,
\label{eq19}
\end{equation}  
where $\Omega_n = \sqrt{ \Omega_h^2 + n \pi \kappa_h 
\omega_o^2}$ for the quantized orbits inside the halo 
($R_n < R_h$), and $\Omega_n = \sqrt{ \Omega_h^2 
- n \pi \kappa_h \omega_o^2 }$ for the quantized orbits 
outside the halo ($R_n > R_h$), with $\Omega_h^2 > 
n \pi \kappa_h \omega_o^2$.

For the orbits inside ('+') the halo, there is no restriction on 
$n$ as when $n \rightarrow \infty$, then $R_n \rightarrow 0$. 
However, for large values of $n$, the orbits become so dense 
that they form a continuum, which is identified with a central 
core of the halo.  In other words, the halo has its central core, 
in which DM particles can move on any orbit.  Since these orbits 
form the continuum, DM particles may easily collide with each 
other.  From a physical point of view, the situation is different 
in the envelope that surrounds the core as all orbits in the 
envelope are quantized and there is a finite number of them
between the core and the edge of the halo; more details can 
be found in Section 5. 

Now, for the orbits outside ('-') the halo, the value of $n$ is 
restricted by the condition $\Omega_h^2 > n \pi \kappa_h 
\omega_o^2$, which means that there is a value of $n$ for
which $\Omega_h^2 \approx n \pi \kappa_h \omega_o^2$;
in this case, when $\Omega_n \rightarrow 0$, then $R_n 
\rightarrow \infty$.  Since this happens for a finite value of 
$n$, the number of quantized orbits outside the halo can 
be estimated by using $n \approx \Omega_h^2 / (\pi 
\kappa_h \omega_o^2)$, with understanding that $n$ 
is the largest integer resulting from this relationship.

The described quantum structure of the halo and its energy 
spectrum resemble an atom with its available energy levels.
In the following, the theory is now applied to a halo with 
constant density, and then to another halo with a given 
density profile $\rho (R)$; the physical pictures of the DM 
halos emerging from these models are discussed.

\subsection{Halos with constant density}

The simplest density profile is $\rho (R) = \rho_o =$ const, 
with $\rho_o$ being specified.  Then, the mass of halo inside
the radius $R$ is given by $M (R) = 4 \pi \rho_o R^3 / 3$, 
and the mass of the entire halo is $M_h =M (R_h) = 4 \pi 
\rho_o R_h^3 / 3$.  Using these masses, the corresponding 
orbital frequencies are: $\Omega_c = 4 \pi G \rho_o / 3$, 
and $\Omega_h = \Omega_c (R_h) = G M_h / R_h^3 = 4 
\pi G \rho_o / 3$, which shows that the orbital frequency 
remains the same at each point inside the halo [20]; 
however, it must be noted that the orbital velocity $v_h 
= R_h \Omega_h$ at the edge of the halo is smaller by 
the factor $\sqrt{2}$ than the escape velocity $v_{esc}$ 
from the halo.  

For this model, the evaluation of the integral $I_c (R_h)$ 
gives the dimensionless constant $C_{\rho} = 1$, which
means that $\kappa_h = \hbar^2 / (G m^3 R_h)$.  By 
applying the theory developed in Section 4.1, it is seen 
that there no quantized orbits inside the halo because 
all orbital frequencies in this model are the same; thus, 
the quantization rules given by Eqs (\ref{eq17}) and 
(\ref{eq18}) with the '+' sign cannot be applied inside 
the halo.  However, the quantization rules with the '-' 
sign can be applied to space that surrounds the halo, 
and they show that if any DM or OM particles are 
present there, then they must be confined to the orbits
that are quantized; the developed theory allows finding
such orbits and their location around the halo after $m$
and $\omega_o$ are either specified (see Section 4.3) 
or determined by future DM experiments. 

\subsection{Halos with linear density profile}

Let a halo of radius $R_{h}$ and total mass $M_{h}$ have 
its density profile given by 
\begin{equation}
\rho (R) = \rho_c \left ( 1 - \frac{R}{R_h} \right )\ ,
\label{eq20}
\end{equation}  
where $\rho_c = 3 M_h / \pi R_h^3$ is the central density of 
the halo; this density profile is much simpler than the so-called 
Einasto or NFW profiles (e.g., [38-41]); nevertheless, the model 
illustrates well all important aspects of the developed quantum 
theory of DM halo. 

The mass distribution $M(R)$ resulting from this density profile 
gives 
\begin{equation}
\Omega_c^2 (R) = 4 \Omega_h^2 \left [ 1 - \frac{3}{4} 
\left ( \frac{R}{R_h} \right ) \right ]\ .
\label{eq21}
\end{equation}  
In the limits of $R = R_h$ and $R = 0$, one obtains 
$\Omega_{c,min}^2 (R_h) = \Omega_h^2$ and 
$\Omega_{c,max}^2 (0) = 4 \Omega_h^2$,
respectively, which represents the range of orbital 
frequencies allowed in this model.

Using Eq. (\ref{eq21}), the integral given by Eq. (\ref{eq14}) 
can be evaluated, and the result is $I_c (R_h) = (5/2) R_h  
\Omega_h^2$, which shows that $C_{\rho} = 5 / 2.$  Then, 
the resulting spectrum of quantized orbits and energies is given 
by Eqs (\ref{eq17}) and (\ref{eq18}), respectively, with the 
dimensionless constant being $\kappa_h = \hbar^2 / (5 G 
m^3 R_h / 2)$.  Using Eq. (\ref {eq19}) and the obtained 
range of orbital frequencies, one finds $R_{max} = R_h$,
which corresponds to $\Omega_{c,min} = \Omega_h$, and  
$R_{min} = 0.7 R_h$, which corresponds to $\Omega_{c,max} 
= 2 \Omega_h$.  Identifying $R_{min}$ as the radius of a core,
in which the density is $\rho_c =$ const, it is seen that 70\% of
of the size of the halo is its core, but the envelope is only 30\%.   

As already shown in Section 4.2, the quantization rules given by
Eqs (\ref{eq17}) and (\ref{eq18}) apply only to the envelope 
(the '+' sign) and to space that surrounds the halo (the '-' sign).
The reason is that the core's density is constant, which means 
that all its orbital frequencies are the same, hence, they cannot 
be quantized.  The highest orbital frequency is at $R_{min}$,
and then its value decreases towards the edge of the halo,
where it becomes $\Omega_h$.  In the vicinity of the halo,
the orbital frequency continues to decrease with distance and
approaches zero when $R \rightarrow \infty$.

To perform some estimates, let the total mass of the halo be 
$M_h \sim 10^{10} M_{sun} \approx 10^{40}$ $kg$, and 
its radius be $R_h \sim 10^{20}$ $m$.  Then, $\Omega_h^2 
\sim 10^{-30}$ $s^{-1}$, which corresponds to the orbital 
velocity at $R = R_h$ to be about $100$ $km$ $s^{-1}$ 
[42,43].  In addition, the dimensionless constant $\kappa_h$ 
can also be calculated by specifying the mass $m$ of the DM 
particles. Taking $m\approx 10 m_p \approx 10^{-26}$ $kg$, 
where $m_p$ is the proton mass, the resulting $\kappa_h \sim 1$; 
however, if $m \approx 100 m_p$, which is the Higgs boson 
mass, then, $\kappa_h \sim 10^{-3}$.  

Since $\omega_o$ and $m$ are currently unknown, one may 
take $\Omega_h / \omega_o \sim 10$ and $\kappa_h \sim 1$, 
which allows obtaining $\Omega_n$ and using Eq. (\ref{eq19})
to find the values of $R_n$ in the envelope as well as in the 
vicinity of the halo.  It is seen that $R_{min}$ in the envelope 
is reached by a finite number of $n$, and that this number 
increases by three orders of magnitude when $\kappa_h 
\sim 10^{-3}$ is used; in both cases, the density of orbits 
near the core increases, and this increase is more prominent 
in the second case.  On the other hand, the number of 
quantized orbits outside the halo can be calculated by using 
$n \approx \Omega_h^2 / (\pi \kappa_h \omega_o^2)$ 
and the above values of $\Omega_h$, $\kappa_h$ and 
$\omega_o$.  For $\kappa_h \sim 1$, one finds $n \sim 
30$, and for $\kappa_h \sim 10^{-3}$, it is $n \sim 3 
\cdot 10^4$, which shows a strong dependence of the 
number of quantized orbits on the value of $\kappa_h$, 
or more precisely, on the value of the currently unknown 
mass of the DM particles. 
 
Finally, it must be pointed out that the size of the core 
and envelope, as well as the number of quantized orbits 
may significantly change when more realistic density 
profiles for DM halos (e.g., [38-41]) are considered, 
and when the values of $m$ and $\omega_o$ for DM 
are experimentally established.

\section{Physical implications and predictions}

The developed quantum theory based on a new asymmetric 
equation is significantly different than the QM based on the 
Schr\"odinger equation.  The main difference is that the 
theory can be applied to a spherical halo of DM particles, 
and it predicts its quantum structure that contains a core, 
which is free of quantum orbits, and a surrounding envelope 
that may contain many quantized orbits.  This shows that 
the halo's quantum structure resembles that known in atoms.

There are two physical constants $\varepsilon_o$ and $\kappa_h$ 
in the theory that play essential roles in establishing the quantum 
structure of the DM halo on its global scale.  One of these constants 
is the quanta of energy $\varepsilon_o = \hbar \omega_o$, which 
is fixed for DM; however, the value of its characteristic frequency 
$\omega_o$ is unknown.  DM particles may absorb and emit 
$\varepsilon_o$, and they may also interact gravitationally with 
each other by exchanging virtual quanta of energy; for this reason, 
$\varepsilon_o$ is called a {\it dark graviton} [20].  Since the theory 
requires that $\omega_o$ = const, DM particles can only emit or 
absorb radiation with this one frequency.  

The other constant $\kappa_h = \hbar^2 / (G m^3 C_{\rho} R_h)$ 
is dimensional, and an interesting combination of the universal 
constants $\hbar$ and $G$, the mass $m$ of a DM particle, the 
density profile constant $C_{\rho}$, and the radius of the DM halo 
$R_h$. The value of $\kappa_h$ directly affects the orbit and energy 
quantization rules given by Eqs (\ref{eq17}) and (\ref{eq18}), which
can also be used to determine the radius $R_n$ of quantized orbits 
(see Eq. \ref{eq19}).  The results presented in Section 4.1 show that 
there is no restriction on $n$ for the orbits inside the halo because 
when $n \rightarrow \infty$, then $R_n \rightarrow 0$.  However,
for large values of $n$, the orbits become so dense that they 
practically form a continuum, which is identified with a core of the 
halo.  The quantized orbits exist in the envelope that surrounds the 
core, and they also exist in space outside of the halo. These different 
parts of halo are now discussed. 

The fact that the orbits form a continuum inside the core means
that the density inside the core is constant, which implies that 
there are no quantized orbits in the core (see Sections 4.2 and
4.3).  As a result, the DM particles in the core move randomly 
and undergo frequent collisions.  There is a narrow region in 
the immediate vicinity of the core in which many orbits may 
exist close to each other, and form a transition between the 
core and envelope.  In this narrow region, the density of orbits 
maybe so high that the DM particles on such orbits may directly 
interact with the randomly moving and colliding particles of the 
core causing some particle's exchange.  Moreover, the core may 
be filled with free dark gravitons (the fixed quanta of energy 
$\varepsilon$ [20]), which may contribute to the gravitational 
wave background [44] and make it different than such wave 
background of the envelope that surrounds this core;  the 
difference may be detected observationally by the NANOGrav 
detector [45] other similar future experiments.

The envelope that surrounds the core in the halo contains the 
quantized orbits, and the DM particles are confined to these 
orbits.  The main reason for this confinment is the lack of 
transitions between the orbits, which is caused by the fact 
that the orbital frequency differences $\Delta \Omega_n = 
\Omega_{n+1} - \Omega_{n}$ are not exact multiples 
of $\varepsilon_o$, because they also depend on $\kappa_h$. 
As a result, DM particles cannot move from one orbit to 
another by emitting or absorbing $\varepsilon_o$.  Instead, 
the particles are confined to their orbits with no other quantum 
restrictions, since they are spinless and have no charge.  An 
exception could be a narrow region very close to the core, 
where many orbits may exit, and the quantum jumps (by 
absorbing or emitting $\varepsilon_o$) between these orbits 
and the core may take place.

In the considered model of halos, all DM particles are inside
the halo with space surrounding the halo being matter free.
Nevertheless, the presence of the halo itself is resposnible 
for the existence of quantized orbits outside the halo.  The 
developed theory demonstrates that the number of such 
orbits is a sensitive function of the two basic parameters 
of the theory, namely, $\omega_o$ and $\kappa_h$, 
which are currently unknown; the smallest are these 
parameters, the largest is the number of quantized orbits
outside the halo.  

It must be pointed out that the predicted quantum structure 
of the halo is only valid for the elementary DM particles, for
which the wave-particle duality is important, and whose 
masses are within the range of the proton and Higgs masses. 
For classical particles, whose masses are many orders of  
magnitude higher, the parameter $\kappa_h$ becomes very 
small and the quantum effects are negligible; this seems to 
be consistent with the fact that DM cannot form gravitationally 
bounded objects larger than pairs of DM particles [20].
Moreover, the theory cannot be applied to OM particles in 
gaseous nebulae because their electromagnetic and plasma 
effects would supersede all gravitational effects described in
this paper for DM halos.\\

The main aim of this paper is to present a quantum theory 
of DM halos and discuss physical implications and predictions
of the theory.  An important topic of the origin of DM is 
outside the scope of the presented approach.  Nevertheless, 
the topic has been considered in the literature and different 
explanations have been proposed  (e.g., [1-8]).  For example,
a relationship between dissipative DM and intermediate black 
holes was suggested [46], and it was also proposed that 
primordial black holes can be responsible for the origin of 
either DM or large-scale cosmic structures, depending on 
their initial masses [47].  Another interesting idea is that 
microscopic warmholes and extra-dimensions can generate 
both DM and dark energy [48].  However, as pointed out 
by these authors, the problem of the origin and physical 
nature of DM remains one of the most challenging and 
unsolved problems in modern cosmology.      

\section{Conclusions}

A quantum theory of DM particles in a spherical halo is 
developed based on a new asymmetric equation [19,20], 
which is complementary to the Schr\"odinger equation.
The two physical parameters that determine the theory 
and its predictions are the quanta of energy $\varepsilon_o 
= \hbar \omega_o$, with $\omega_o$ being a fixed 
frequency, and a new dimensional parameter $\kappa_h 
= \hbar^2 / (G m^3 C_{\rho} R_h)$, which combines the 
universal constants $\hbar$ and $G$, the mass $m$ of DM 
particles, the density profile constant $C_{\rho}$, and the 
radius of the DM halo $R_h$. 

The theory predicts that the halo contains a core that 
is surrounded by an envelope.  The core is filled with 
free and randomly moving DM particles that collide 
with each other.  The envelope contains the quanitized 
orbits on the halo's global scale with the lowest frequency 
(energy) orbit located at the edge of the halo.  The orbital 
frequency (energy) inreases towards the core and the 
quantized orbits become more densely populated to 
form a continuum inside the core.  The described 
quantum structure of the halo resembles an atom.

To determine the distribution of quantized orbits in 
the envelope, the density profile in the halo as well 
as its radius and total mass must be specified.  This
means that the distribution of quantized orbits is 
consistent with the global physical parameters of 
the halo, and that the orbits are populated by DM 
particles, so that the density profile is accounted for.
Since DM particles that populate the orbits are spinless
and have no charge, there no are other quantum limits 
on the number of particles on each orbit.

DM particles are allowed to emit or absorb the quanta
of energy, $\varepsilon_o$, called dark graviton.  
However, since differences between the quantized 
orbits are not exact multiples of $\varepsilon_o$, as 
they also depend on $\kappa_h$, DM particles are 
permanently confined to their orbits.  An exception 
could be a narrow region in the immediate vicinity of 
the core, where many orbits may exit.  DM particles 
may undergo quantum jumps between these orbits 
and the core by absorbing or emitting $\varepsilon_o$,
which requires that the differences between some 
orbits and the core are exactly equal to the quanta 
of energy $\varepsilon_o$.  The existence of the
dark gravitons, or a sea of these gravitons in the 
core, may cause its gravitational wave background 
to be different than that generated by the envelope 
of the halo.\\

\bigskip\noindent
{\bf Acknowledgment:} 
The author is indebted to an anonymous referee for providing 
comments and suggestions that allowed significantly improved 
this paper.  The author also thanks Dora Musielak for valuable 
comments on the earlier version of this manuscript.  This 
work was partially supported by Alexander von Humboldt 
Foundation.\\


\noindent
{\bf References}

\begin{thebibliography}{}

\bibitem{1} M.J. Rees, Dark Matter - Introduction, Astro-Physics 361 (2003) 2427 
\bibitem{2} K. Freeman, and G. McNamara, In Search of Dark Matter, Springer, 
                  Praxis, Chichester, 2006
\bibitem{3} L. Papantonopoulos, L. (Editor), The Invisible Universe: Dark Matter and
                    Dark Energy, Lecture Notes in Physics 720, Springer, Berlin – Heidelberg, 2007
\bibitem{4} R.H. Sanders, The Dark Matter Problem: A Historical Perspective, Cambridge 
                    Uni. Press, Cambridge, 2010
\bibitem{5} J.A. Frieman, M.B. Turner, and D. Huterer, Ann. Rev. Astron. Astrophys. 
                  46 (2008) 385
\bibitem{6} G. Bartone, and D. Hooper, Rev. Mod. Phys. 90 (2018) 045002
\bibitem{7} E. Oks, New Astron. Rev. 93 (2021) 101632
\bibitem{8} L. Hui, Ann. Rev. Astron. Astrophys. 59 (2021) 247
\bibitem{9} E. Aprile et al., Phys. Rev. Let. 122 (2019) 141301
\bibitem{10} J.M. Overduin, and P.S. Wesson, Phys. Rep. 283 (2004) 337
\bibitem{11} R. Barbier et al., Phys. Rep. 420 (2005) 1
\bibitem{12} K. Sugita, Y. Okamoto, M. Sekine, Int. J. Theor. Phys. 47 (2008) 2875
\bibitem{13} N. Arkani-Hamed, D.P. Finkbeiner, T.R. Slatyer, and N. Weiner, Phys. 
                    Rev. D. 79 (2009) 015014
\bibitem{14} E. Komatsu et al., ApJS 192 (2011) 18
\bibitem{15} T.M. Undagoita, and L. Rauch, arXiv:1509.08767v1 [physics.ins-det] 26 Sep 2015 
\bibitem{16} S. Giagu, Front. Phys. 7 (2019) 75
\bibitem{17} Y.J. Ko, and H. K. Park, arXiv:2105.11109v3 [hep-ph] 4 June 2021
\bibitem{18} T.B. Watson, and Z.E. Musielak, Int. J. Mod. Phys. A, 35 (2020) 2050189 (10pp)
\bibitem{19} Z.E. Musielak, Int. J. Mod. Phys. A, 36 (2021) 2150042 (12pp)
\bibitem{20} Z.E. Musielak, Int. J. Mod. Phys. A, 37 (2022) 2250137 (10pp)
\bibitem{21} M. Ackermann et al., Phys. Rev. Let. 107 (2011) 241302
\bibitem{22} A. Ibarra, D. Tran and C. Weniger, Int. J. Mod. Phys., 28 (2013)
                   1330040 (48pp)
\bibitem{23} T.M. Undagoita and L. Rauch,  arXiv:1509.08767v1 [physics.ins-det]
                   26 Sep 2015	
\bibitem{24} Y. Hochberg, Y.F. Kahn, R.K. Leane, et al., Nature Rev. Phys. 4 (2022), 637
\bibitem{25} N. Aghanim, et al., Astron. Astrophys. 641 (2020) A6 (67 pages)
\bibitem{26} E. Merzbacher, Quantum Mechanics, Wiley \& Sons, Inc., New York, 1998
\bibitem{27} E.P. Wigner, Ann. Math. 40 (1939) 149 
\bibitem{28} V. Bargmann, Ann. Math. 59 (1954) 1
\bibitem{29} J.-M. Levy-Leblond, Comm. Math. Phys. 6 (1967) 286
\bibitem{30} J.-M. Levy-Leblond, J. Math. Phys. 12 (1969) 64
\bibitem{31} Y.S. Kim and M.E. Noz, Theory and Applications of the Poincar\'e Group, 
                    Reidel, Dordrecht, 1986
\bibitem{32} E. Schr\"odinger, Ann. d. Physik 79 (1926) 361 
\bibitem{33} S.-J. Sin, Phys. Rev. D 50 (1994) 365
\bibitem{34} W. Hu, R. Barkana, and A. Gruzinov, Phys. Rev. Lett. 85 (2000) 1158
\bibitem{35} S.C. Spivey, Z.E. Musielak, and J.L. Fry, MNRAS 428 (2013) 712 
\bibitem{36} S.C. Spivey, Z.E. Musielak, and J.L. Fry, MNRAS 448 (2015) 1574 
\bibitem{37} Z.E. Musielak, Adv. Math. Phys. Vol. 2023 (2023) Article ID 5736419 (11 pages)
\bibitem{38} J. Einasto, and U. Haud, Astron. Astrophys. 223 (1989) 89 
\bibitem{39} J.F. Navarro, C.S. Frenk, and S.D. White,  Astrophys. J. 
                   462 (1996) 563 
\bibitem{40} D. Merritt, et al., Astron. J. 132 (2006) 2685
\bibitem{41} J.F. Navarro, et al., MNRAS 402 (2010) 21
\bibitem{42} A.N. Bushev, MNRAS 417 (2011) L83
\bibitem{43} K. Garrett and G. Duda, Adv. Astron. Vol. 2011 (2011) Article ID 968283 (22 pages)
\bibitem{44} J.D. Romano, and N.J. Cornish, Living Rev. Relativ. 20 (2017) (1): 2
\bibitem{45} G. Agazie, A. Anumarlapudi, A.M. Archibald, et al., Astrophys. J. Let.
                   951 (2023) L40
\bibitem{46} G. D'Amico, P. Panci, A. Lupi, S. Bovino, and J. Silk, MNRAS 473 (2018) 328 
\bibitem{47} B. Carr, and J. Silk , MNRAS 478 (2018) 3756 
\bibitem{48} A.R. El-Nabulsi, Mod. Phys. Let. 36 (2021) 2150042
\end{thebibliography}
\end{document}